\documentstyle[twoside,epsfig]{article}

\catcode`\@=11
\long\def\@makefntext#1{
\protect\noindent \hbox to 3.2pt {\hskip-.9pt  
$^{{\eightrm\@thefnmark}}$\hfil}#1\hfill}		

\def\@makefnmark{\hbox to 0pt{$^{\@thefnmark}$\hss}}	
	
\def\ps@myheadings{\let\@mkboth\@gobbletwo
\def\@oddhead{\hbox{}
\rightmark\hfil\eightrm\thepage}   
\def\@oddfoot{}\def\@evenhead{\eightrm\thepage\hfil
\leftmark\hbox{}}\def\@evenfoot{}
\def\sectionmark##1{}\def\subsectionmark##1{}}

\oddsidemargin=\evensidemargin
\newcounter{sectionc}\newcounter{subsectionc}\newcounter{subsubsectionc}
\renewcommand{\section}[1] {\vspace{12pt}\addtocounter{sectionc}{1} 
\setcounter{subsectionc}{0}\setcounter{subsubsectionc}{0}\noindent 
	{\tenbf\thesectionc. #1}\par\vspace{5pt}}
\renewcommand{\subsection}[1] {\vspace{12pt}\addtocounter{subsectionc}{1} 
	\setcounter{subsubsectionc}{0}\noindent 
	{\bf\thesectionc.\thesubsectionc. {\kern1pt \bfit #1}}\par\vspace{5pt}}
\renewcommand{\subsubsection}[1] {\vspace{12pt}\addtocounter{subsubsectionc}{1}
	\noindent{\tenrm\thesectionc.\thesubsectionc.\thesubsubsectionc.
	{\kern1pt \tenit #1}}\par\vspace{5pt}}
\newcommand{\nonumsection}[1] {\vspace{12pt}\noindent{\tenbf #1}
	\par\vspace{5pt}}

\newcounter{appendixc}
\newcounter{subappendixc}[appendixc]
\newcounter{subsubappendixc}[subappendixc]
\renewcommand{\thesubappendixc}{\Alph{appendixc}.\arabic{subappendixc}}
\renewcommand{\thesubsubappendixc}
	{\Alph{appendixc}.\arabic{subappendixc}.\arabic{subsubappendixc}}

\renewcommand{\appendix}[1] {\vspace{12pt}
        \refstepcounter{appendixc}
        \setcounter{figure}{0}
        \setcounter{table}{0}
        \setcounter{lemma}{0}
        \setcounter{theorem}{0}
        \setcounter{corollary}{0}
        \setcounter{definition}{0}
        \setcounter{equation}{0}
        \renewcommand{\thefigure}{\Alph{appendixc}.\arabic{figure}}
        \renewcommand{\thetable}{\Alph{appendixc}.\arabic{table}}
        \renewcommand{\theappendixc}{\Alph{appendixc}}
        \renewcommand{\thelemma}{\Alph{appendixc}.\arabic{lemma}}
        \renewcommand{\thetheorem}{\Alph{appendixc}.\arabic{theorem}}
        \renewcommand{\thedefinition}{\Alph{appendixc}.\arabic{definition}}
        \renewcommand{\thecorollary}{\Alph{appendixc}.\arabic{corollary}}
        \noindent{\tenbf Appendix \theappendixc #1}\par\vspace{5pt}}
\newcommand{\subappendix}[1] {\vspace{12pt}
        \refstepcounter{subappendixc}
        \noindent{\bf Appendix \thesubappendixc. {\kern1pt \bfit #1}}
	\par\vspace{5pt}}
\newcommand{\subsubappendix}[1] {\vspace{12pt}
        \refstepcounter{subsubappendixc}
        \noindent{\rm Appendix \thesubsubappendixc. {\kern1pt \tenit #1}}
	\par\vspace{5pt}}

\newcommand{\textlineskip}{\baselineskip=13pt}
\newcommand{\smalllineskip}{\baselineskip=10pt}

\def\eightcirc{
\begin{picture}(0,0)
\put(4.4,1.8){\circle{6.5}}
\end{picture}}
\def\eightcopyright{\eightcirc\kern2.7pt\hbox{\eightrm c}} 

\newcommand{\copyrightheading}[1]
	{\vspace*{-2.5cm}\smalllineskip{\flushleft
	 }}

\newcommand{\publisher}[2]{{\begin{center}\footnotesize\smalllineskip 
	Received #1\\
	Revised #2
	\end{center}
	}}

\def\abstracts#1#2#3{{
	\centering{\begin{minipage}{4.5in}\footnotesize\baselineskip=10pt
	\parindent=0pt #1\par 
	\parindent=15pt #2\par
	\parindent=15pt #3
	\end{minipage}}\par}} 

\def\keywords#1{{
	\centering{\begin{minipage}{4.5in}\footnotesize\baselineskip=10pt
	{\footnotesize\it Keywords}\/: #1
	\end{minipage}}\par}}

\renewenvironment{thebibliography}[1]
        {\frenchspacing
	 \ninerm\baselineskip=11pt
         \begin{list}{\arabic{enumi}.}
        {\usecounter{enumi}\setlength{\parsep}{0pt}     
	 \setlength{\leftmargin 12.7pt}{\rightmargin 0pt} 
         \setlength{\itemsep}{0pt} \settowidth
	{\labelwidth}{#1.}\sloppy}}{\end{list}}

\newcounter{itemlistc}
\newcounter{romanlistc}
\newcounter{alphlistc}
\newcounter{arabiclistc}
\newenvironment{itemlist}
    	{\setcounter{itemlistc}{0}
	 \begin{list}{$\bullet$}
	{\usecounter{itemlistc}
	 \setlength{\parsep}{0pt}
	 \setlength{\itemsep}{0pt}}}{\end{list}}

\newcommand{\fcaption}[1]{
        \refstepcounter{figure}
        \setbox\@tempboxa = \hbox{\footnotesize Fig.~\thefigure. #1}
        \ifdim \wd\@tempboxa > 5in
           {\begin{center}
        \parbox{5in}{\footnotesize\smalllineskip Fig.~\thefigure. #1}
            \end{center}}
        \else
             {\begin{center}
             {\footnotesize Fig.~\thefigure. #1}
              \end{center}}
        \fi}

\newcommand{\tcaption}[1]{
        \refstepcounter{table}
        \setbox\@tempboxa = \hbox{\footnotesize Table~\thetable. #1}
        \ifdim \wd\@tempboxa > 5in
           {\begin{center}
        \parbox{5in}{\footnotesize\smalllineskip Table~\thetable. #1}
            \end{center}}
        \else
             {\begin{center}
             {\footnotesize Table~\thetable. #1}
              \end{center}}
        \fi}

\def\@citex[#1]#2{\if@filesw\immediate\write\@auxout
	{\string\citation{#2}}\fi
\def\@citea{}\@cite{\@for\@citeb:=#2\do
	{\@citea\def\@citea{,}\@ifundefined
	{b@\@citeb}{{\bf ?}\@warning
	{Citation `\@citeb' on page \thepage \space undefined}}
	{\csname b@\@citeb\endcsname}}}{#1}}

\newif\if@cghi
\def\cite{\@cghitrue\@ifnextchar [{\@tempswatrue
	\@citex}{\@tempswafalse\@citex[]}}
\def\citelow{\@cghifalse\@ifnextchar [{\@tempswatrue
	\@citex}{\@tempswafalse\@citex[]}}
\def\@cite#1#2{{$\null^{#1}$\if@tempswa\typeout
	{IJCGA warning: optional citation argument 
	ignored: `#2'} \fi}}

\def\pmb#1{\setbox0=\hbox{#1}
	\kern-.025em\copy0\kern-\wd0
	\kern.05em\copy0\kern-\wd0
	\kern-.025em\raise.0433em\box0}

\def\fnt#1#2{\footnotetext{\kern-.3em
	{$^{\mbox{\scriptsize #1}}$}{#2}}}

\def\ps@myheadings{%
    \let\@oddfoot\@empty\let\@evenfoot\@empty
    \def\@evenhead{\slshape\leftmark\hfil}
    \def\@oddhead{\hfil{\slshape\rightmark}}
    \let\@mkboth\@gobbletwo
    \let\sectionmark\@gobble
    \let\subsectionmark\@gobble
    }
\font\tenrm=cmr10
\font\tenit=cmti10 
\font\tenbf=cmbx10
\font\bfit=cmbxti10 at 10pt
\font\ninerm=cmr9

\font\eightrm=cmr8

\def\qed{\hbox{${\vcenter{\vbox{		    
   \hrule height 0.4pt\hbox{\vrule width 0.4pt height 6pt
   \kern5pt\vrule width 0.4pt}\hrule height 0.4pt}}}$}}

\def\bsc{{\sc a\kern-6.4pt\sc a\kern-6.4pt\sc a}}  	
\def\bflatex{\bf L\kern-.30em\raise.3ex\hbox{\bsc}\kern-.14em 
T\kern-.1667em\lower.7ex\hbox{E}\kern-.125em X} 

\begin{document}
\setlength{\textheight}{7.7truein}  

\thispagestyle{empty}

\markboth{\protect{\footnotesize\it Opinion formation models based on game theory.}}
{\protect{\footnotesize\it }}

\normalsize\textlineskip

\setcounter{page}{1}

\copyrightheading{}			

\vspace*{0.88truein}

\centerline{\bf Opinion formation models based on game theory}
\centerline{\bf }

\vspace*{0.37truein}

\centerline{\footnotesize Alessandro Di Mare}
\centerline{\footnotesize\it Scuola Superiore di Catania}
\baselineskip=10pt
\centerline{\footnotesize\it Catania, I-95123 , Italy }
\centerline{\footnotesize\it E-mail: aldimare@ssc.unict.it}

\vspace*{15pt}          
\centerline{\footnotesize Vito Latora}
\centerline{\footnotesize\it Dipartimento di Fisica e Astronomia, 
Universit\'a di  Catania, and}
\centerline{\footnotesize\it  INFN, Sezione di Catania}
\baselineskip=10pt
\centerline{\footnotesize\it Catania, I-95123 , Italy }
\centerline{\footnotesize\it E-mail: latora@ct.infn.it}

\vspace*{0.225truein}
\publisher{(received date)}{(revised date)}

\vspace*{0.25truein}
\abstracts{A way to simulate the basic interactions between two individuals 
with different opinions, in the context of strategic game theory,  
is proposed. Various games are considered, which produce different kinds  
of opinion formation dynamics. 
First, by assuming that all individuals (players) are equals, 
we obtain the bounded confidence model of continuous opinion 
dynamics proposed by Deffuant et al. 
In such a model a tolerance threshold is defined, 
such that individuals with difference in opinion larger than 
the threshold can not interact.
Then, we consider that the individuals have 
different inclinations to change opinion and different abilities in 
convincing the others. 
In this way, we obtain the so-called ``Stubborn individuals and Orators'' (SO) 
model, a generalization of the Deffuant et al. model, 
in which the threshold tolerance is different for every couple 
of individuals. We explore, by numerical simulations, 
the dynamics of the SO model, and we propose further 
generalizations that can be implemented.  
}{}{}

\vspace*{5pt}
\keywords{Sociophysics; opinion dynamics; game theory.}

\vspace*{1pt}\textlineskip	
\section{Introduction}		
\vspace*{-0.5pt}
\noindent

The last years have seen an increasing interest in the physics 
community for the description and modeling of social systems. 
In particular, Monte Carlo simulations have become an important part 
of {\it sociophysics} \cite{weidlich,staufrev1},
enlarging the field of interdisciplinary applications 
of statistical physics. 
Most of the sociophysics models, such as that by Deffuant et al., 
that by Hegselmann-Krause, and the Sznajd model \cite{staufrev2,santorev}, 
dealing with 
opinion dynamics and consensus formation, have the limit of considering
that the individuals in a society are all equals. 
Conversely, an important feature of 
any real system, to be considered in the modeling 
of social system, is the presence of individuals with different 
inclinations to change idea \cite{pluchino}, 
as well as individuals with different abilities in convincing the others.  
\\
In this paper we focus on the model proposed by Deffuant et al.~\cite{Deff}, 
showing how the standard version of the model 
can be derived from basic principles in the framework of {\em game theory}. 
Moreover, in the context of game theory, the model can be easily 
generalized in different directions to take into account 
of the presence of individuals with different characteristics. 
In particular, we show how the introduction of a distribution 
of the individual inclinations to change and of the 
ability to convince the others, that produces what we have named the 
{\em Stubborn Individuals and Orators (SO) model}, 
can affect the opinion dynamics of a social group. 
\\
The paper is organized as follows. In Section 2 we briefly review 
the standard sociophysics models, in particular  
the model by Deffuant et al. In Section 3 we propose 
a way to derive models of opinion dynamics in 
the framework of game theory. 
The method we propose is based on a set of basic assumptions 
on the characteristics of the individuals (the players of the 
game) and on the payoffs for each of the possibile actions,  
and on the idea of Nash 
equilibrium for games with perfect information. 
We show how some simple models, including the model 
by Deffuant et al. can be derived. We then 
consider the SO model, 
a generalized model considering agents with 
different ability in convincing the others and with 
different inertia in changing ideas. 
In Section 4 we explore the dynamics of the SO model by means 
of extensive numerical simulations.   
In Section 5 we draw the conclusions and we 
outline further possible generalizations and future 
developments.

\vspace*{1pt}\textlineskip	
\section{Opinion formation models}	
\vspace*{-0.5pt}
\noindent
The standard models of opinions dynamics ~\cite{staufrev1,staufrev2,santorev,report} 
in sociophysics deal with $N$ {\em individuals} or {\em agents}. 
Each individual $i=1,...,N$ is characterized, at time $t$, by 
an opinion $S_i (t)$.  
The opinions can be integer numbers (for instance +1 or -1) as in the 
Sznajd model \cite{Sznajd}, or real numbers in the range $[0,1]$ as in the 
model by Deffuant et al. \cite{Deff} and in the Hegselmann and Krause model 
\cite{HK}. Each agent is in continuous interaction with the 
other agents. The opinion of an agent changes under the influence of the 
other individuals according to very simple deterministic rules. 
For instance, in the Sznajd model on a two-dimensional square lattice, 
at each time step, two  randomly selected neighboring agents 
transfer their opinion to the six neighbors if and only if 
the two agents of the pair share the same opinion. 
In the {\em model by Deffuant et al.}~\cite{Deff}, at each time step $t$,  
two randomly selected neighboring agents $i$ and $j$ check their opinions 
$S_i(t)$ and $S_j(t)$ to see whether an exchange of opinion is possible. 
If the two opinions differ by more than a fixed threshold
parameter $\epsilon$ ($0<\epsilon<1$), called the {\em confidence bound},
both opinions remain unchanged. 
If, instead, $| S_i(t) - S_j(t)| < \epsilon$, then each opinion moves
in the direction of the other as: 
\begin{equation}
\left\{ \begin{array}{c}
	 S_i(t+1) = S_i(t) + \mu  [ S_j(t) - S_i(t) ]  \\
	 S_j(t+1) = S_j(t) - \mu  [ S_j(t) - S_i(t) ]  \\
       \end{array}
\right. 
\end{equation} 
with $\mu$ being a second tunable parameter ($0< \mu \le 1/2$). 
In the basic model, the threshold $\epsilon$ is taken as 
fixed in time and constant across the whole population. In particular, 
the value $\mu=1/2$ corresponds to the case in which the two 
opinions take their average $[S_i(t) - S_j(t)] /2$  \cite{Deff}.

To see if and how a consensus emerges out of initially different opinions,
the models are usually started with a random initial distribution of
opinions.
The dynamics is followed until the system reaches an equilibrium state
characterized by the existence of one or several opinion groups,
according to the value of the control parameters of the models.
For instance, the basic Sznajd model with random sequential updating
always leads to a consensus on a regular lattice of any dimension $D$
(and even if more than two opinions are allowed).
In particular, 
one observes a phase transition as
a function of the initial concentration $p$ of up spins.
If $p=0.5$, then, at the end of the dynamics, half of the samples will
have $S_i=+1~\forall i$ and the remaining half will have
$S_i=-1~\forall i$.
For $p < 1/2$ all samples end up with  $S_i=-1 ~\forall i$, while for
for $p > 1/2$ they all end up in the other attractive fixed
point $S_i=+1 ~\forall i$ \cite{staufrev1}.
In the Deffuant model with $\mu=0.5$ instead, all opinions converge to a single 
central one for $\epsilon > 1/2$, while for $\epsilon < 1/2$ different opinions
survive, with a number of surviving opinions that varies as
$1/\epsilon$, as also confirmed by analytical arguments \cite{bennaim}.

\vspace*{1pt}\textlineskip	
\section{Game theory and generalized models of opinion formation}	
\vspace*{-0.5pt}
\noindent

The main idea behind any of the previous models is to simulate how the 
opinions change in time by analyzing the very basic facts, 
that is: two individuals with different opinions on 
a given topic meet and discuss, trying to convince each other,  
or to find somehow a certain agreement about the topic. Of course,  
it is not obvious that the two individuals do find a common agreement on the 
topic, this depending basically on the specific characteristics of the 
two individuals (some of the individuals in a real social system are easy 
to convince, other are less flexible, some are good orators or distinguished 
for skills and power in convincing the others, while some others 
are timid and reserved), and also on some external factors 
(the time length and the strength of the interaction, the pressure of 
the external environment or of the dominant ideas and fashions).  

In this paper we propose to examine the basic interactions between the 
two individuals within the framework of game theory \cite{osborne}. 
For this reason, from now on, the two individuals will be also 
referred to as the two players. 
In particular, we make use of the concept of   
Nash equilibrium for games with perfect (or complete) information. 

In general, a {\em strategic game} is a model of interacting decision makers. 
It consists of: 
\begin{itemize}
 \item a set of $N$ {\em players} or {\em decision-makers};
 \item for each player $i$ ($i=1,...,N$), a set of possible 
       {\em actions} $A=\{a, b, c, ...\}$;  
 \item for each player, {\em preferences} over the set of action profiles 
 (i.e. the list of all the players' actions).
\end{itemize}
One way to describe the player's preferences is to specify for each possible 
pair of actions, the action the player prefers, or to note that the player 
is indifferent between the actions. Alternatively one can represent the 
preferences by a {\em payoff function}, which associates a number with
each action, in such a way that actions with higher numbers are preferred. 
More precisely, the payoff function $u$ represents a player's preferences 
if, for any couple of actions $a$ and $b$ in $A$, $u(a) > u(b)$ if and only 
if the player prefers $a$ to $b$. 
A simple example can be that of a person that is faced with three 
vacation packages, to New York, Paris and Venice. She prefers the package 
to Venice to the other two, which she regards as equivalent. Her 
preferences can be represented by any payoff function that assigns 
the same number to New York and Paris, and a higher number to Venice. 
For example, we can set $u(a)=u(b)=0$ and $u(c)=1$, where 
$a, b, c$ represent, respectively, the three packages. 
The fundamental hypothesis in game theory is that each player tries  
to maximize her benefit. This is usually called the hypothesis 
of {\em rational choice}, and means that, in any given situation, the 
decision-maker chooses the member of the available subset of $A$ that is best 
according to her preferences. 
Moreover, the strategic games considered here deal with situations in 
which actions are chosen once and for all (whereas there are games, 
named extensive games, allowing for the possibility that plans may be revised 
as they are carried out \cite{osborne}).  
In the example above, the decision-maker will decide to go to Venice. 
In this simple example, we have only one decision-maker choosing 
an action from a set $A$, and caring only about this action. 
In the general case (that is of interest in this article), some 
of the variables that affect a player are the actions of other 
decision-makers, so that the decision-making problem is more challenging 
than that of an isolated player. The typical example is firms selling an 
item and competing for business. Each firm controls its price, but not the 
other firms's prices. Each firm cares however, about all the firms 
prices, because these prices affects its sales. How should a firm choose 
its prices in such a case ? 
In this case, the best action for any given player depends in general 
on the other players' actions. So when choosing an action, a player must
have in mind the actions the other players will choose. 
That is, she must form a belief about the other players' actions. 
On what basis can such a belief be formed ? We consider here games 
in which each player's belief is derived from her past experience playing 
the game, and this experience is sufficiently extensive that she knows 
how her opponents will behave. No one tells her the actions the 
opponents will choose, but her previous involvement in the game 
leads her to be sure of these actions. 
These are called {\em games with complete 
information}, since in such games each player knows all the details of 
the game and of its elements. 

In summary, in the strategic games we consider, there are two 
different components. First, each 
player chooses her action according to the model of rational choice, 
given her belief about the other players' actions. 
Second, every player's belief about the other players' actions is correct. 
These two components are embodied in the following definition of 
{\em Nash equilibrium} for such games \cite{osborne,nash}: 
\begin{itemize} 

\item  A Nash equilibrium is an action profile 
$a^* \equiv (a^*_1, a^*_2,...,a^*_N$) - where $a^*_1$ denotes the 
action chosen by player 1,  $a^*_2$ the action chosen by player 2 and so on-,  
with the property that no player $i$ can do better by choosing an action 
different from $a^*_i$, given that every other player $j$ adheres to $a^*_j$.    
\end{itemize}

This definition implies neither that a strategic game necessarily 
has a Nash equilibrium, nor that it has at most one. In general, 
some games have a single Nash equilibrium, some possess no Nash 
equilibrium and others have many Nash equilibria. 
A Nash equilibrium corresponds to a ``steady state'' 
of the system: if, whenever the game is played, the action profile is the 
Nash equilibrium $a^*$, then no player has a reason to choose any 
action different from her component of $a^*$. 
In practice, there is no pressure on the action profile to change. 
Expressed differently, a Nash equilibrium embodies a stable ``social norm'': 
if everyone else adheres to it, no individual wishes to deviate from 
it. The second component of the theory of Nash equilibrium 
(that the players' beliefs about each other's actions are correct) implies, 
in particular, that two players' beliefs about a third player's action are the 
same. For this reason, the condition is sometimes said to be that the 
players' ``expectations are coordinated'' \cite{osborne}. 

The situations to which we wish to apply the theory of Nash equilibrium 
is the process of decision making in the formation of an opinion. 
We simulate the elementary interaction between individuals in a society 
by means of a strategic game. 
Then, we get a model of opinion formation, by iterating 
the game many times, i.e. by choosing at each time step a group of 
individuals and allowing them to play the game. In particular 
we assume that each game is played by only two players ($N=2$), since here we 
limit to the particular case in which the dynamics of opinion formation 
is based on the continuous interaction between couples of individuals. 
(This is not always true. There are many real situations in which 
the elementary process of opinion formation is based on the 
mutual interaction of groups of more than two individuals. 
Nevertheless, our ideas can be generalized to games 
with $N>2$).   
We suggest a list of different possible games, the difference being in:   
\begin{enumerate}
\item the number and kind of actions that one individual can choose from;
\item the characteristics of the two individuals.   
\end{enumerate}
For instance, in the simplest model (defined in Subsection 3.1),  
we assume that the individuals playing the game are all equals  
and can choose between two possibilities, either 
to mantain or to change their opinion. 
In the following, more complex models, we consider more than two actions 
for each player to choose from (e.g. introducing the possibility that 
the two players find an agreement). Moreover, we introduce a way to take 
into account that in a social group there are individual with different 
skills and abilities.

Before moving to the descriptions of the models and their equilibrium 
we want to stress that interactions in real social systems do not in general 
correspond exactly to the idealized setting described above 
(rational choice and complete information).  
For example, in some cases, the players do not have much experience with 
the game. In some other cases it could be useful to introduce non-rational 
players.  
Whether or not the notion of Nash equilibrium is appropriate in any given 
situation is a matter of judgment. 
In some cases, a poor fit with the idealized setting 
may be mitigated by other considerations. For example, inexperienced players 
may be able to draw conclusions about their opponents' likely actions 
from their experience in other situations, or from other sources. 
Ultimately, the test of the appropriateness of the notion of Nash
equilibrium is whether it gives us insights into the problem at hand, 
that is to develop models of opinion formation.

\subsection{Game I}
\noindent
In the most basic case, each of the two players of the game 
(named, from now on, player A and player B) can choose between two different 
actions: to mantain or to change opinion. 
As in any game we need to fix the actions' payoffs $u$.  
The payoff, for a player, is the function of her and the other 
player' actions. 
Of course, each player wants to convince the other one that her opinion 
is correct; on the other hand she does not want to accept easily the 
other player' opinion. Therefore, for each player, we fix the following 
payoffs: 
\begin{itemlist}
 \item u=+a if the other player changes her opinion  
 \item u=+b if the player keeps her opinion 
 \item u=-b  if the other player keeps her opinion 
 \item u=-a if the player changes her opinion
\end{itemlist}
where $a,b \in \Re$ and $0<b<a$. We take $b<a$, since 
we assume that a player gets the greatest satisfaction when 
is able to convince the opponent. Of course, this is just an 
hypothesis so that, in principle, also the choice $b>a$ would 
be an equally valid possibility. More in general, we should consider 
four different numbers for the payoffs, respectively: 
$a>0$, $b>0$, $c<0$ and $d<0$. 
Here, for the sake of simplicity, we assume  $c=-b$ and $d=-a$. 
In table \ref{table1}, we report the payoff $u$ for players A and B, 
for each of the strategies (action profiles).  
\begin{table}[htbp] 
\tcaption{Game I: payoffs' tables for player A and player B}
\centerline{\footnotesize PAYOFFS FOR A}
\centerline{\footnotesize\smalllineskip
\begin{tabular}{l c c }\\
\hline
{} &B changes & B keeps\\
\hline
 A changes &0    &-a-b \\
 A keeps   &+a+b &0 \\
\hline\\
\end{tabular}}
\centerline{\footnotesize PAYOFFS FOR B}
\centerline{\footnotesize\smalllineskip
\begin{tabular}{l c c }\\
\hline
{} &B changes & B keeps\\
\hline
 A changes &0    &a+b \\
 A keeps   &-a-b &0 \\
\hline\\
\end{tabular}}
\label{table1}
\end{table}
The two tables can be thought as two matrices $M^A$, $M^B$, 
whose entry $m_{ij}$ represents the payoff, respectively for players A 
and B, when A chooses the strategy $i$ and B chooses the strategy $j$. 
E.g., $m^A_{12}$ ($m^B_{12}$) is the payoff 
for player A  (player B)
when A chooses the action {\em change} and B chooses the action  
{\em keep}. Such a payoff is obtained by considering that player A is in 
the following condition: 
she changes her opinion, while the opponent mantains her opinion, 
therefore $m^A_{12} = -a -b$.  
On the other hand, player B mantains her opinion while 
the opponent changes her opinion, so that $m^B_{12} = +a+b$. 
With the same method we obtain the payoff for all the situations.
The two matrices are: 
\begin{equation}
M^A=\left( \begin{array}{cc}
	0    & -a-b  \\
	+a+b & 0     \\
       \end{array}
\right) 
~~~~~~
M^B=\left( \begin{array}{cc}
	0    & a+b   \\
	-a-b & 0     \\
       \end{array}
\right)
\nonumber
\end{equation} 
After creating the game, we want to foresee the actions that will 
be taken by the two players. It is easy to prove that the game has 
a single Nash equilibrium in (2,2), i.e.  
when both players choose the strategy {\em keep}. 
In fact, if we fix that B chooses keep, then the player A can choose 
between a negative payoff $-a-b$ or 0. 
Therefore she chooses the payoff 0, that corresponds to the strategy 
keep. 
In the same way, the player B chooses keep if we fix that A 
chooses keep. 
Therefore, neither player gets a greater payoff modifying her 
strategy, if the other player does not change her own. 
Consequently, two players of game I will always mantainin 
their own opinion, never reaching an agreement. In conclusion, 
a model of opinion dynamics, in which at each time step a 
couple of individuals is chosen at random among the $N$ individuals 
and play game I, will produce no time evolution: every individual 
will mantain the initial opinion.

\subsection{Game II}
\noindent
We now introduce a new possibility: the {\it agreement}. This means 
that the two players can decide to change their own 
opinion with an intermediate one (not the initial opinions, nor the 
intermediate one needs to be better specified here).  
At this point we need to fix the payoff for the two new 
possible actions, considering that a player gets a certain 
satisfaction if she is able to shift the opponent's opinion 
to an intermediate one. We fix the following two new payoffs:
\begin{itemlist}
 \item  u=+c if the other player changes her opinion with an 
intermediate one   
 \item u=-c if the player changes her opinion with an intermediate one 
\end{itemlist}
where $c \in \Re$ and $c<a$. 
In table \ref{table2} we represent the payoffs for the various 
action profiles. 
\begin{table}[htbp] 
\tcaption{Game II: payoffs' tables for player A and player B}
\centerline{\footnotesize PAYOFFS FOR A}
\centerline{\footnotesize\smalllineskip
\begin{tabular}{l c c c}\\
\hline
{}         &  B changes & B keeps & B agrees \\
\hline
 A changes &0    &-a-b  & -a+c\\
 A keeps   &+a+b &0     & +b+c\\
 A agrees  &+a-c &-b-c  & 0   \\
\hline\\
\end{tabular}}
\centerline{\footnotesize PAYOFFS FOR B}
\centerline{\footnotesize\smalllineskip
\begin{tabular}{l c c c}\\
\hline
{}         &  B changes & B keeps & B agrees  \\
\hline
 A changes &0    &+a+b  & +a-c\\
 A keeps   &-a-b &0     & -b-c\\
 A agrees  &-a+c &+b+c  & 0   \\
\hline\\
\end{tabular}}
\label{table2}
\end{table}
The two matrices for the new conflicting opinion game are:  
\begin{equation}
M^A=\left( \begin{array}{ccc}
	0    & -a-b  & -a+c\\
	+a+b & 0     & +b+c\\
        +a-c & -b-c  & 0   \\
       \end{array}
\right) 
~~~
M^B=\left( \begin{array}{ccc}
	0    & a+b  & a-c  \\
	-a-b & 0    & -b-c \\
        -a+c & +b+c & 0    \\
       \end{array}
\right)
\nonumber
\end{equation} 
Notice that in this game, as in the previous one, the two 
matrices are trivially related since $M^A= -M^B$. 
The game has a single Nash equilibrium in the point $(2,2)$. 
This means that, although the two players have, in principle, also 
the possibility of finding an agreement, they choose the strategy 
of mantaining their own ideas. Finally, the outcome of a model 
of opinion dynamics based on game II, would not produce 
results different from the trivial results of the model based on game I, 
i.e. no dynamics at all.

\subsection{Game III}
\noindent
In the two previous games we have not taken into account 
the distances between the two players' opinions.  
Obviously, two individuals with close enough opinions can reach  
easier the agreement than two people with very different 
opinions. This has led to the introduction of the confidence 
bound mechanism in the model by Deffuant et al. (see Section 2). 
Hence, it can be useful also here, in the context of strategic games,  
to introduce a distance $d$ between the opinions of the players, 
and a corrective $d$-dependent term that makes the agreement 
easier in the game when $d$ is smaller. 
We fix the following payoffs:  
\begin{itemlist}
 \item u=+a if the other player changes her opinion  
 \item u=+b if the player keeps her opinion 
 \item u=-b if the other player keeps her opinion 
 \item u=-a if the player changes her opinion 
 \item u=+c+1/d if the other player changes her opinion with an 
   intermediate one.   
 \item u=-c+1/d if the player changes her opinion with an intermediate one 
\end{itemlist}
where $d \in \Re$, and $d>0$. 
Notice that, if the two conditions $c + \frac{1}{d} > a$ and 
$-c + \frac{1}{d} > a$ are valid, then both players get a greater 
payoff by choosing the strategy to agree.  
The two conditions are both verified if $d < \frac{1}{a+c}$. 
The payoffs' matrices for the game are: 
\begin{equation}
M^A=\left( \begin{array}{ccc}
	0                & -a-b              & -a+c+\frac{1}{d}\\
	+a+b             & 0                 & +b+c+\frac{1}{d}\\
        +a-c +\frac{1}{d}& -b-c +\frac{1}{d} &  \frac{2}{d}    \\
       \end{array}
\right)
\nonumber
\end{equation} 
\\ 
~~~
\begin{equation}
M^B=\left( \begin{array}{ccc}
	0               & a+b              & a-c+\frac{1}{d}  \\
	-a-b            & 0                & -b-c+\frac{1}{d} \\
        -a+c+\frac{1}{d}& +b+c+\frac{1}{d} & \frac{2}{d}   \\
       \end{array}
\right)
\nonumber
\end{equation} 
\\
It is easy to prove that this game has two different equilibrium 
points. In fact: 

when $d \le \frac{1}{b+c}$, there is a Nash equilibrium in (3,3).

when $d \ge \frac{1}{b+c}$, there is a Nash equilibrium in (2,2).
\\
If we define the confidence bound $\epsilon$ as the following 
function of the game parameters: 
\begin{equation}
  \epsilon = \frac{1}{b+c} 
  \label{epsilonIII} 
\end{equation} 
an opinion model based on $N$ individuals playing in randomly chosen 
couples game III,  
coincides exactly with the model by Deffuant et al. (with $\mu =0.5$) 
\cite{Deff}, discussed in Section 2. 
In fact, we can assume that the opinions are real numbers in the 
range $[0,1]$, as in the model by Deffuant et al., and we can 
start the $N$ individuals with a uniform random distribution of opinions. 
We then fix the three parameters $a, b, c$, that is equivalent 
to fixing a value of the confidence bound $\epsilon$ (a tuning of the 
parameters $a, b, c$ allows to have $\epsilon$ varying in 
the range $[0,1]$), while the distance $d$ depends on the two players' 
opinions and changes each time the game is played.  
Finally, the resulting model is the following. 
At each time step, two randomly chosen individuals 
play game III. When the distance $d$ between the two opinions 
is smaller than the confidence bound in Equation (\ref{epsilonIII}), 
then the two individual shift their opinions to the average one otherwise 
they keep their own opinions. This is nothing else than the 
model by Deffuant et al.

\subsection{Game IV: The SO model}
\noindent
In the previous models, the individuals are considered all equals. 
They can have different opinions, but they have the same way to interact 
(i.e. in our framework, to play the game).  
Actually, this is far from being true in any real case. What makes 
the world interesting is the diversity of characters and behaviours  
we encounter. In particular, in a real social system, 
there are {\em stubborn} individual, i.e. individuals that do not change 
their own opinion easily, as well as people that change their opinion 
very easily. 
Moreover, there are {\em orators}, i.e. individuals with a 
certain influence in group processes and a well known ability in 
convincing the others \cite{centrality},  
as well as individuals that are not good in convincing the others. 
In order to take this into account in our models of opinion 
formation, we introduce two new variables, 
so that every individuals in our model is characterized by two values, 
the former representing the oratory ability and the latter representing 
the stubbornness of a person. We assume that each player 
is characterized by a couple of real numbers $p,q \in ]0,1[$ 
(the so-called {\em characteristic parameters} of the individual), 
where the first variable, $p$, represents the probability for a 
player to convince the opponent, and the second value, $q$, 
is the probability that a player keeps her own opinion. 
Considering, as in the previous cases, a game with only two players, 
A and B, we assume that 
$p_A$, $q_A$ and $p_B$, $q_B$ are the characteristic parameters of the two 
players. 
Obviously, in the new game, the payoffs for each player should also depend  
on the four numbers $p_A, q_A, p_B, q_B$. This can be easily understood 
in the following way. Let us consider the action ``A convinces B'' 
and suppose to iterate the game $n$ times (with $n \ll 1$). 
In $n$ steps, the expected total payoff of A is $n a P$, where 
$a$ is the payoff for the same action in the previous game, and 
$P$ is the probability that A convinces B. 
The latter is the product of $p_A$, the probability that A is able 
to convince, and $1-q_B$, that is the probability that B 
let himself to be convinced. 
Therefore, for every step A has a mean payoff equals to 
$a p_A (1-q_B)$. In a similar way we obtains the payoffs for all 
the other actions. Furthermore, we consider a distance-dependent 
term as in model III: since the average value of 
the payoffs is one fourth of that in the previous game, 
we choose $\frac{1}{4d}$ as corrective term. 
Finally, the two matrices representing the payoffs for A and B are the 
following:  
\begin{equation}
M^A=\left( \begin{array}{ccc}
	a(P-X)                & -aX-bY              & cP-aX+\frac{1}{4d}\\
	&&\\
	bQ+aP           & b(Q-Y)               & bQ+cP+\frac{1}{4d}\\
	&&\\
  aP-cX+\frac{1}{4d}   & -cX-bY+\frac{1}{4d} & c(P-X)+\frac{1}{2d}\\
       \end{array}
\right) 
\nonumber
\end{equation} 
\\ 
~~~
\begin{equation}
M^B=\left( \begin{array}{ccc}
	a(X-P)                & aX+bY               & aX-cP+\frac{1}{4d}\\
	&&\\
	-bQ-aP               & b(Y-Q)               & -bQ-cP+\frac{1}{4d}\\
	&&\\
  cX-aP+\frac{1}{4d}   & cX+bY+\frac{1}{4d}   & c(X-P)+\frac{1}{2d}\\
       \end{array}
\right) 
\nonumber
\end{equation} 
where we have set $P=p_A(1- q_B)$, $Q=q_A(1-p_B)$, $X=p_B (1-q_A)$ and $Y= q_B (1-p_A)$. 
The Nash equilibrium of the game changes as a function of the distance $d$: 
\\ 
if $d \le D(p_B,q_A)$ and  $d \le D(p_A,q_B)$, then there is a Nash equilibrium in (3,3);
\\
if $d \ge D(p_B,q_A)$ and  $d \ge D(p_A,q_B)$, then there is a Nash equilibrium in (2,2);
\\
if $d \ge D(p_B,q_A)$ and  $d \le D(p_A,q_B)$, then there is a Nash equilibrium in (2,3);
\\
if $d \le D(p_B,q_A)$ and  $d \ge D(p_A,q_B)$, then there is a Nash equilibrium in (3,2); 
\\
where we have defined the two following functions: 
\begin{eqnarray}
D(p_B,q_A) = \frac{1}{4 [b q_A+c p_B-(b+c)q_A p_B]} \nonumber
\\
D(p_A,q_B) = \frac{1}{4 [b q_B+ c p_A - (b+c)p_A q_B]}
\label{ede}
\end{eqnarray}

The game we have proposed has three main differences with respect to 
the basic interaction mechanism in the model by Deffuant et al.:

\begin{enumerate} 

\item It is possible that a player chooses the strategy ``agreement'' 
while the other player chooses the strategy ``keep''. 
This is not possible in the Deffuant et al.~model.

\item The largest distance that allows for the agreement  
(confidence bound) depends on the characteristic parameters 
of the couple of players. 

\item The shifting of the two players is in general 
different, so that it is necessary to introduce two different 
variables, $\mu$ and $\nu$ (one for each player),  
to indicate how much the two player shift their opinions. 

\end{enumerate}

If we indicate by $S_A(t)$ and $S_B(t)$ the two opinions 
at the time $t$, we can write the time evolution as:  
\begin{equation}
\left\{ \begin{array}{c}
	 S_A(t+1) = S_A(t) + \mu  [ S_B(t) - S_A(t) ]  \\
	 S_j(t+1) = S_B(t) - \nu  [ S_B(t) - S_A(t) ]  \\
       \end{array}
\right. 
~~~~~
\left\{ \begin{array}{c}
	 0 < \mu < 0.5      \\
	 0 < \nu < 0.5      \\
       \end{array}
\right. 
\nonumber
\label{final_g4}
\end{equation} 
where $\mu$ is a function of $p_A$ and $q_B$, while $\nu$ is a function 
of $p_B$ and $q_A$. To determine the functions $\mu(p_A,q_B)$ and 
$\nu(p_B,q_A)$ we notice that they should obey to the following obvious 
conditions: 
\begin{equation}
 \label{}
 \frac{\partial \mu}{\partial p_A} >0; ~~~
 \frac{\partial \mu}{\partial q_B} <0; ~~~
 \frac{\partial \nu}{\partial p_B} >0; ~~~
 \frac{\partial \nu}{\partial q_A} <0; ~~~~~~~
 \forall p_B, q_B, p_A, q_A \in ]0,1[
\end{equation}
For instance, the first two conditions state that $\mu(p_A,q_B)$ should be 
an increasing function of $p_A$ and a decreasing function of $q_B$. 
One possibility is to choose the two following functions: 
\begin{equation}
\label{munu}
\mu(p_A,q_B)= \frac{p_A (1-q_B)}{2}  ~~~~~~  \nu(p_B,q_A)= \frac{p_B (1-q_A)}{2}
\end{equation}       
Such a choice is in perfect agreement with the contraints $0<\mu <0.5$ and $0<\nu <0.5$. 
Finally, we note that, if  $d \le D(p_A,q_B)$, then player B chooses the action 
``agreement'' whatever is the strategy choosen by A. 
Analogously,  if  $d \le D(p_B,q_A)$, then player A chooses to agree 
whatever is the strategy choosen by B. 
Therefore, the strategic game we have proposed can be resumed in 
the definition of the two following functions: 
\begin{equation}
\mu(p_A,q_B) =\left\{ \begin{array}{c}
	 \frac{p_A (1-q_B)}{2} \\
	 0                     \\
       \end{array}
     \right. 
 \begin{array}{c}
	 if~ d \le D(p_A,q_B) \\
	 if~ d >   D(p_A,q_B) \\
       \end{array} 
\label{mu_game4}
\end{equation}
\begin{equation}
\nu(p_B,q_A) = \left\{ \begin{array}{c}
	 \frac{p_B (1-q_A)}{2}\\
	 0                    \\ 
       \end{array}
\right. 
 \begin{array}{c}
	 if~ d \le D(p_B,q_A) \\
	 if~ d >   D(p_B,q_A) \\
       \end{array} 
\label{nu_game4}
\end{equation} 
In conclusion the ``stubborn individuals and orators'' (SO) model we 
propose, based on the strategic game IV, is the following. 
The model is fixed by choosing the values of the three control parameters 
$a, b, c$ (that is equivalent to choose a value of the confidence 
bound $\epsilon$ in the Deffuant et al model). 
The distance $d$ depends on the two players' opinions and changes 
each time the game is played.  
The $N$ individuals are, as usually, started with a 
random distribution of opinions in the range $[0,1]$. Moreover, each 
player $i$ ($i=1,2,...N$) is now characterized by two variables, 
the characteristic parameters $p_i$ and $q_i$, distributed among the 
players according to two given distribution functions: $F_1(p)$ and 
$F_2(q)$.  
At each time step, two individuals, let say A and B, are chosen at random 
and interact by playing the game. The results of the game depends on the 
distance $d$ between the two players' opinions and on the 
characteristic parameters $p_A, q_A, p_B, q_B$. The two players' opinions 
after the interaction (i.e. after the game) are shifted according to 
Equations (\ref{final_g4}), where the values of $\mu(p_A,q_B)$ and 
$\nu(p_B,q_A)$ are calculated through Equations 
(\ref{mu_game4}) and (\ref{nu_game4}),  and 
through the functions in Equation (\ref{ede}). In particular, 
notice that, when $ d >   D(p_A,q_B)$,  $\mu$ is equal to zero and the opinion 
of player A remains unchanged: $S_A(t+1)= S_A(t)$. Analogously, 
when $ d >   D(p_B,q_A)$, $\nu$ is equal to zero and $S_B(t+1)= S_B(t)$. 
Finally, the opinion dynamics model (the SO model) based on game IV, 
consists in iterating this procedure at each time steps.

\vspace*{1pt}\textlineskip	
\section{Montecarlo simulations of the SO model}		
\vspace*{-0.5pt}
\noindent

In this Section we turn our attention to the numerical simulation of 
the dynamics of the SO model. 
As shown in the previous section, the model depends on three different 
parameters $a$, $b$, and $c$. The three parameters are not independent. 
In fact $a$ does not appear explicitly in Equations (\ref{ede}), and 
plays the role of a normalization parameter, being  
only necessary to fix the maximum value of $b$ and $c$ 
(since we have assumed that $b<a$ and $c<a$. 
Consequently, the number of parameters can be reduced by fixing 
the value of $a$ and by considering the two normalized parameters: 
\begin{equation}
\beta=b/a 
~~~~~~~
\gamma=c/a
\end{equation}
By definition, we have  $0< \beta <1$ and $0< \gamma <1$. 
From Equations (\ref{munu}), we can notice that the mean value of $\mu$ 
and of $\nu$ is equal to $<\mu>=<\nu>=\frac{1}{8}$. 
Hence, if we suppose that the satisfaction of a player is a 
linear function of the shifting, then we can assume $\gamma=\frac{1}{8}$. 
In this way, the only independent parameter is $\beta$, since there 
are no reasons to fix a particular value for the satisfacion of a player 
to mantain the opinion.

Finally, in the numerical simulation we investigate the behaviour of the model 
for $\gamma=\frac{1}{8}$ and for different values of $\beta$. 
In particular, we consider a population of $N$ agents ($N>>1$)  
with an initial (at time $t=0$) heterogeneous distribution of opinions (every 
opinion in the range [0,1] being equally probable). Each agent is described  
by two characteristic parameters $p$ and $q$. We assume that $p$ and $q$ 
are uniformly distributed in the range $]0,1[$, i.e. that 
the two distribution functions $F_1(p)$ and $F_2(q)$ are equal to a 
constant. 
We evolve the system supposing that each player has the same 
probability to interact with any other (a more realistic possibility, not 
considered here, would be that of imagining the individuals interacting on 
a complex topology \cite{report}. 
As explained in Section 3.4, at each time step, two randomly chosen agents 
try to convince each other of their opinions: they decide whether to change 
their opinion on the basis of the rules of game IV. 

\begin{figure}[htbp] 
\vspace*{13pt}
\centerline{\psfig{file=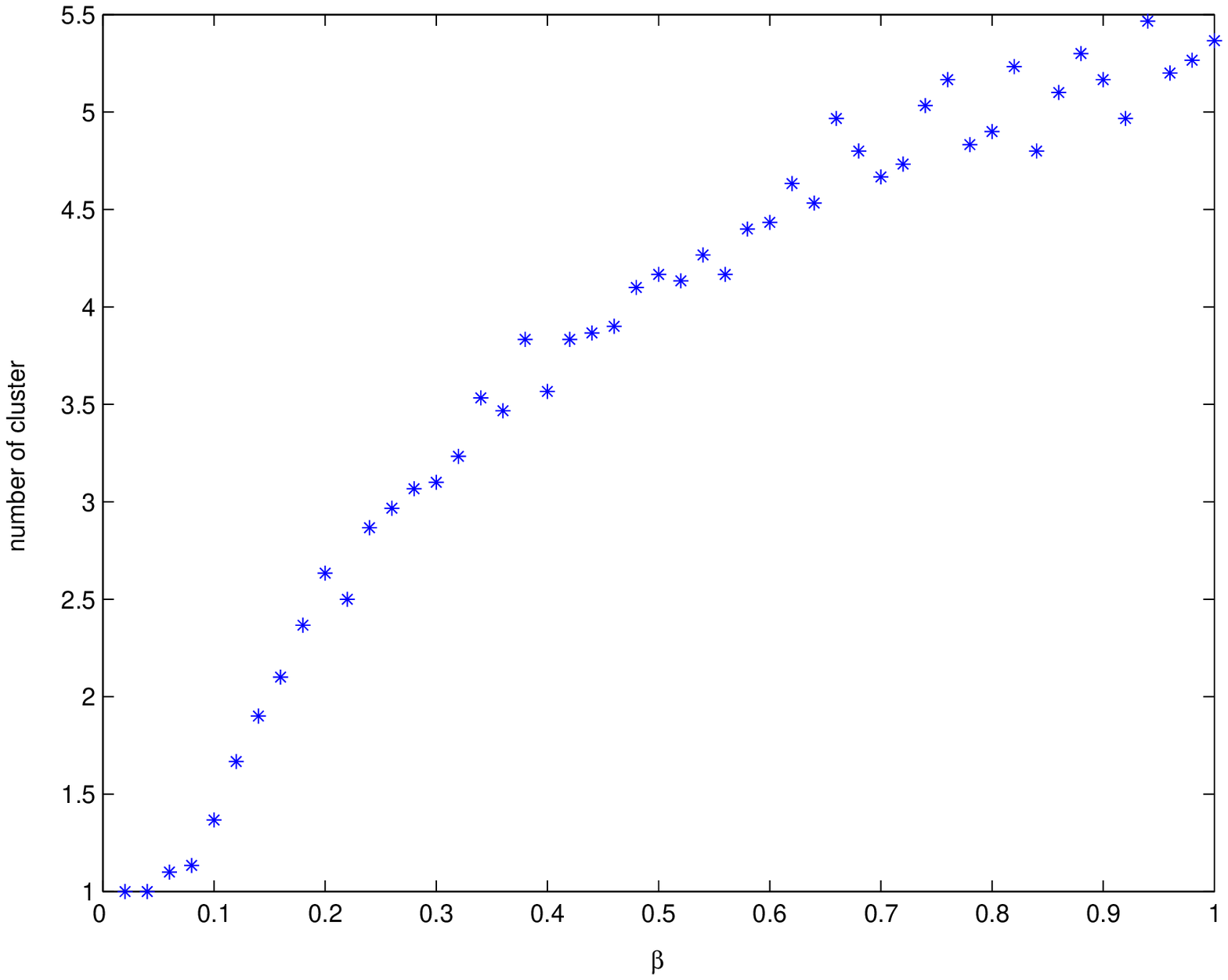,width=4.TRUEIN}} 
\vspace*{13pt}
\fcaption{SO model with $N=500$, $\gamma=1/8$, $\beta$ variable, and a  
uniform distribution of initial opinions and of the individual characteristic 
parameters $p$ and $q$. 
The number of clusters with a size larger than $0.1 N$ at time $t_f= 5 \times 10^5$
is shown as a function of the parameter $\beta$.}
\label{fig1}
\end{figure}

\begin{figure}[htbp] 
\vspace*{5pt}
\centerline{\psfig{file=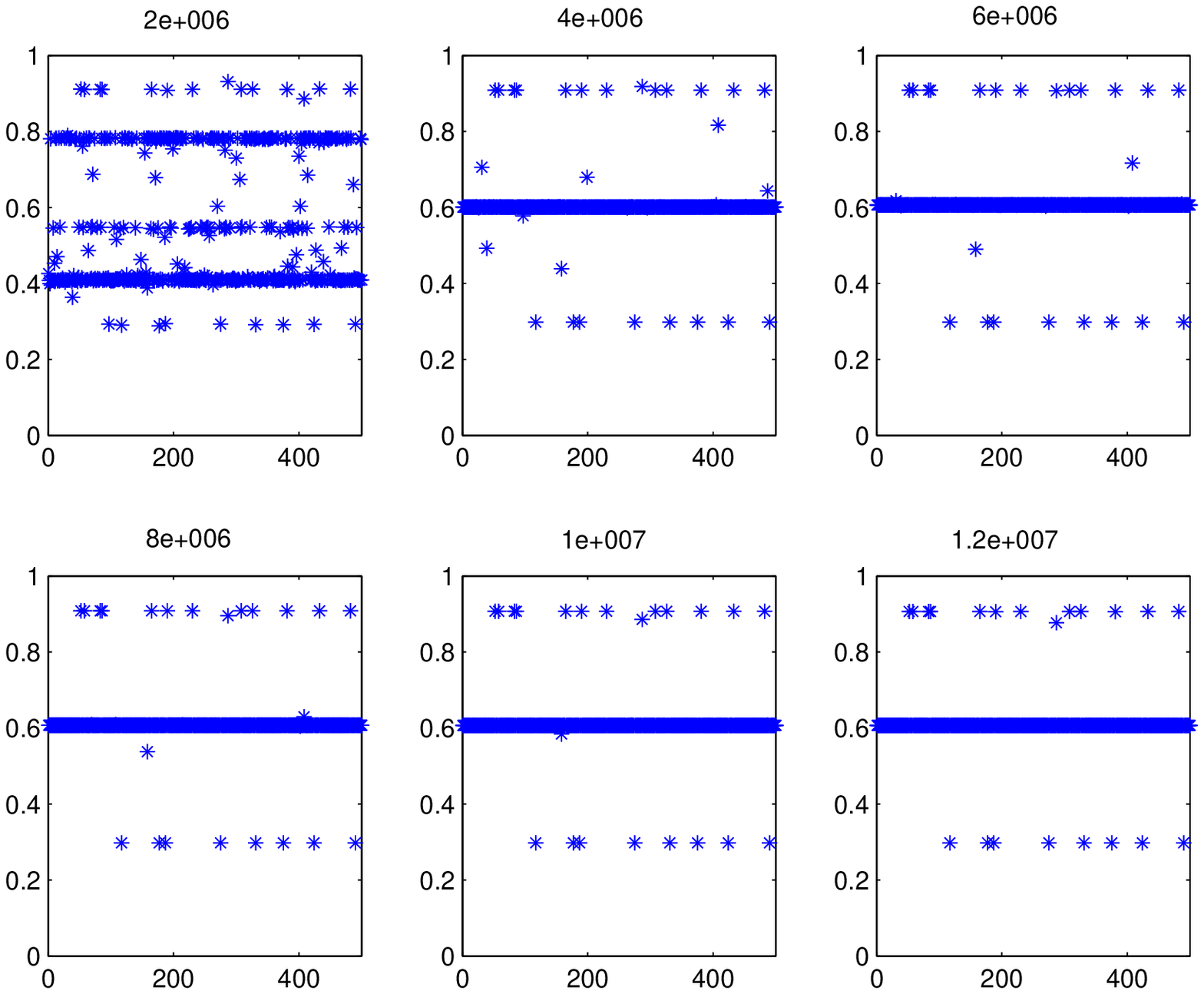,width=4.TRUEIN}} 
\vspace*{5pt}
\fcaption{SO model with $N=500$, $\gamma=1/8$, $\beta=0.8$, 
a uniform distribution of initial opinions and a uniform distribution 
of the individual characteristic parameters $p$ and $q$. 
We report the distribution of opinions among the 
500 agents at six different times: 
$t_1=2 \times 10^6$, $t_2=4 \times 10^6$, $t_3=6 \times 10^6$, 
$t_4=8 \times 10^6$, $t_5=1 \times 10^7$, $t_6=1.2 \times 10^7$. }
\label{fig2}
\end{figure}

In Fig.~\ref{fig1}, we show the number of large clusters, $N_c$,  
obtained at a fixed time $t_{f}$, as a function of $\beta$. 
A cluster is a group of people that share the same opinion. 
In the figure we consider only large clusters, i.e. clusters whose size is 
larger that $10\%$ of the total population.  
We have considered $N = 500$ and $t_{f}= 5 \times 10^5 $ time steps. 
Each of the points reported in the figure has been obtained as an average over 
30 different realizations for the same value of $\beta$.  
%
We observe that $N_c$ is an increasing function of $\beta$. This is an  
obvious consequence of the model: in fact, $\beta$ represents the payoff 
(normalized to $a$) of a player if she mantains her opinion. Hence, an increasing 
value of $\beta$ causes in a player a natural inclination to find the agreement 
only with players with similar opinions, and therefore an increasing number of cluster. 
On the contrary, for small $\beta$, the players tend to change their opinions 
creating immediately a small number of clusters. 
As a further step, we have checked numerically whether the asymptotic state 
of the opinion distribution changes as a function of $\beta$.  
The simulations confirm the tendency of the system to reach a final equlibrium 
with a single large cluster of opinions for any value of $\beta$ in [0,1].
The time to reach the equilibrium depends strongly from $\beta$ and less from 
the different realizations of initial conditions. 
In the Fig.~\ref{fig2} we show the typical dynamical evolution for the 
case $\beta=0.8$. In each panel we plot the agents' opinions for six 
different times. 
%
For $t> 5 \times 10^6$ we notice the presence of a few small groups of opinions 
and a single large cluster 
containing about 96\% of the total population,  
Similar results are obtained for other realizations of the initial conditions and 
for different values of $\beta$. 

The presence of a single large cluster, observed at large times in the simulations,  
is a consequence of the {\it all to all} interactions (any couple of players 
is allowed to interact in this version of the model).  
If we analyze the first of equations (\ref{ede}), 
we notice that $D(p_B,q_A)$ takes only positive values because: 
\begin{equation}
	bq_A+cp_B-(b+c)q_A p_B>0 \ \ \Leftrightarrow\ \ \frac{b}{p_B}+\frac{c}{q_A}>b+c
\end{equation}
is true $\forall p_B, q_A\in\left(0,1\right)$
Analogously, $D(p_A,q_B)$ is positive for any value of $p_A$ and $q_B$. 
This implies the existence, for any choice of the parameters, of a positive distance 
of opinions for which the agreemeent is possible.  
When we start the agents with a heterogeneous opinion distribution,   
for any given player $i$, it is certainly possible to find another 
player that can find an agreement with $i$.  
If we iterate the game IV for a long time, then there is a large probability that 
all the players reach a common agreement. 
Furthermore, if we consider an agent with an opinion next to $1$, she 
has a larger probability to play with another agent with a opinion lower 
than her. 
The same reasoning can be applied to an agent with an opinion close to $0$. 
Consequently, players with extreme opinions moves to central opinions and, 
after a large time we observe a single large cluster with  
a central opinion (e.g. at $S \approx 0.6$ in Fig.~\ref{fig2}). 
An exception to this behaviour are those players who avoid the agreement 
with most of the other players. E.g. let us consider a player with  
$q=\frac{1}{2}$. It can be easily proven that such a player   
chooses the strategy {\it to agree} only when the distance with the opponent's 
opinion is very small. In fact, in the best case, i.e. when the opponent is an 
extremely good oratory characterized by $p=1$, a simple calculation gives 
$D(p,q)=0.08$. 
Consequently, the distance $d$ between the two individuals should be smaller 
than 0.08 for these two individuals to find a common agreement. 
If the dynamical evolution of opinions leads to isolate a player with 
$q=\frac{1}{2}$ at a distance larger that 0.08 from all the other players, such 
a player will never move from her opionion. 
This is the reason why, after a long time, we observe some (in general very few) 
isolated small clusters.

The tendency towards the formation of a single large cluster 
of opinions can be quantified, at each time step, by the calculation 
of the Gini coefficient of the opinion distribution.  
The {\em Gini coefficient} $G$ is a measure commonly used in economics and ecology 
to describe inequalities in the distribution of resource 
in a population \cite{dagum,lorenz}. 
In order to calculate the Gini coefficient of a generic empiric distribution one 
has first to computer the {\em Lorenz curve} of the distribution. 
In our case, the Lorenz curve of the opinion distribution is 
obtained in the following way. 
%
\begin{figure}[htbp] 
\vspace*{13pt}
\centerline{\psfig{file=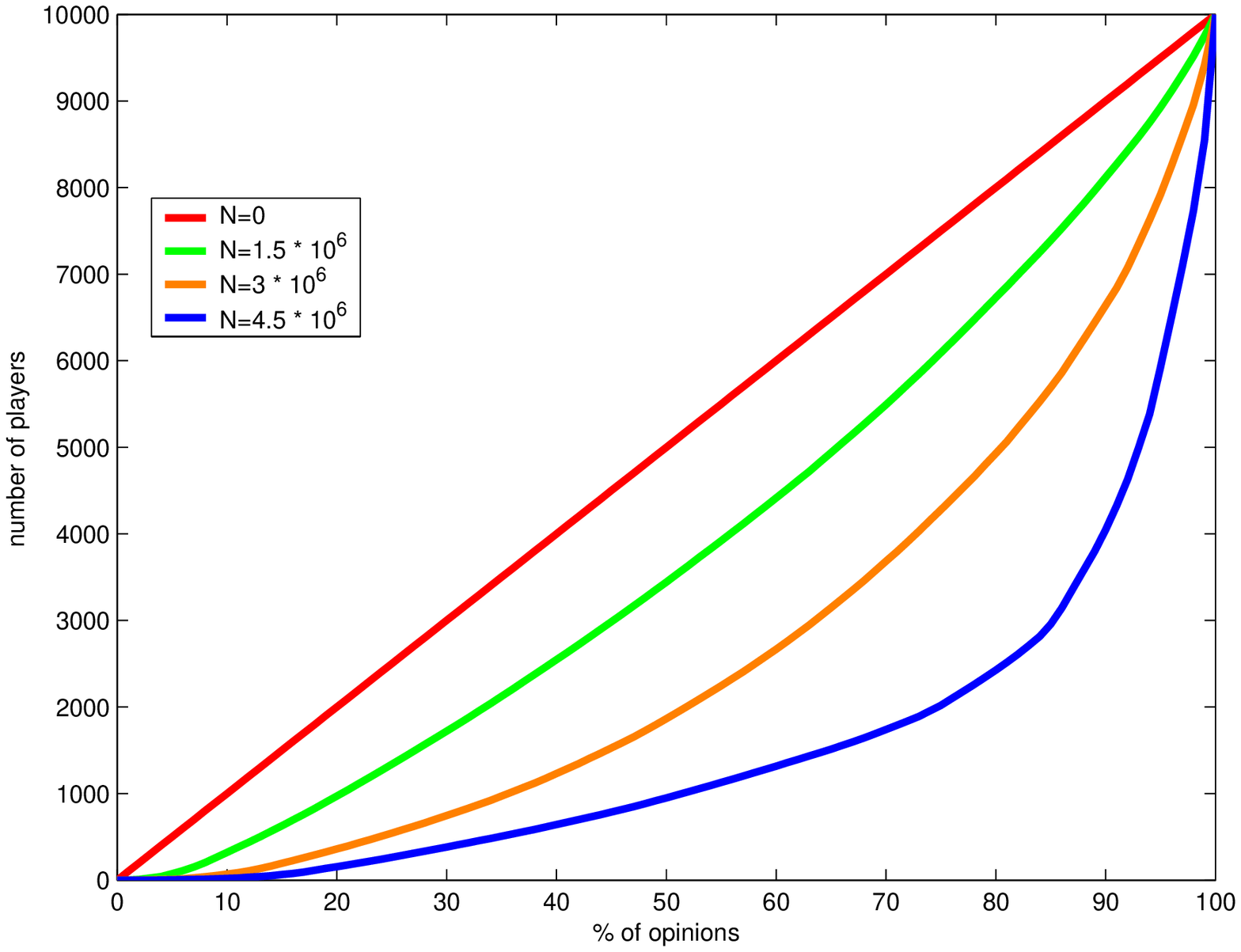,width=3.5TRUEIN}} 
\vspace*{13pt}
\fcaption{SO model with $N=10000$, $\gamma=1/8$, $\beta=0.8$ and a 
uniform distribution of initial opinions and of the individual characteristic 
parameters $p$ and $q$. 
Lorenz curves of the opinion distribution. The number 
of players is reported, 
as a function of the percentage of opinions (see text for details),  
for four different times: $t_1=0$, $t_2=1.5 \times 10^6$, $t_3=3 \times 10^6$ and 
$t_4=4.5 \times 10^6$. 
}
\label{fig3}
\end{figure}
We divide the opinion range [0,1] in $M$ intervals (classes) of size $\Delta s$. 
Class $m$, with $m=1,...M$, contains $n_m$ individuals, namely those having an opinion 
in the range $[(m-1) \Delta s, m  \Delta s]$, with the normalization 
$\sum_{m=1}^{M} n_m = N$. The importance (richness) of a class is measured by 
the number of individuals it contains: the reachest class is the one containing the 
largest number of individuals.  
We then sort the classes in increasing order of $n_m$ (starting from the classes 
with the smallest number of individuals, up to the richest ones). 
Finally, in Fig.~\ref{fig3}) we report (on the y-axis) 
the percentage of individuals,  
as a function of the percentage of the classes considered,  
in increasing order of importance (on the x-axis). 
This is the Lorenz curve of the opinion distribution. We name such a function $y=F(x)$. 
In particular, in  Fig.~\ref{fig3}) we consider the result of a 
simulation of the SO model with $N=10000$ individuals, and we report the Lorenz 
curves $F(x)$ obtained at four different times: 
$t_1=0$, $t_2=1.5 \times 10^6$, $t_3=3 \times 10^6$ and $t_4=4.5 \times 10^6$. 
Notice that, at time $t_1=0$, we have a uniform distribution of 
opinions and, as expected, the Lorenz curve $y=F(x)$ coincides with the 
line of perfect equality $y=x$. For larger times, the Lorenz curve tends more and 
more towards the Lorenz curve for the most heterogeneous distribution, that is   
$F(x)=0$ for $x\in [0,1[$, and $F(x)=1$ for $x=1$. 
This is eventually obtained when all the opinions are in the same class, so 
that the opinion classe are inequally populated: all of them are empty but a 
single one containing all the individuals. 
 The Gini coefficient, $G$, can be calculated by comparing the Lorenz curve 
$y=F(x)$ of the distribution, with the line of perfect equality $y=x$ (the Lorenz 
curve of a perfectly homogeneous distribution). $G$ is defined graphically as 
the ratio of two surfaces: the area between the line of perfect equality and the Lorenz 
curve, and the area between the line of perfect equality and the line 
of perfect inequality. One gets:   
\begin{equation}
  G = 1 - 2 \int^{1}_{0} F(x)dx
\end{equation}
By definition, the coefficient $G$ ranges from a minimum value of zero, in the case 
in which all the classes are equivalently occupied, 
to a maximum value of 1 in a population in which every class 
except one has a size of zero, i.e. all the individuals share the same opinion. 
%
\begin{figure}[htbp] 
\vspace*{8pt}
\centerline{\psfig{file=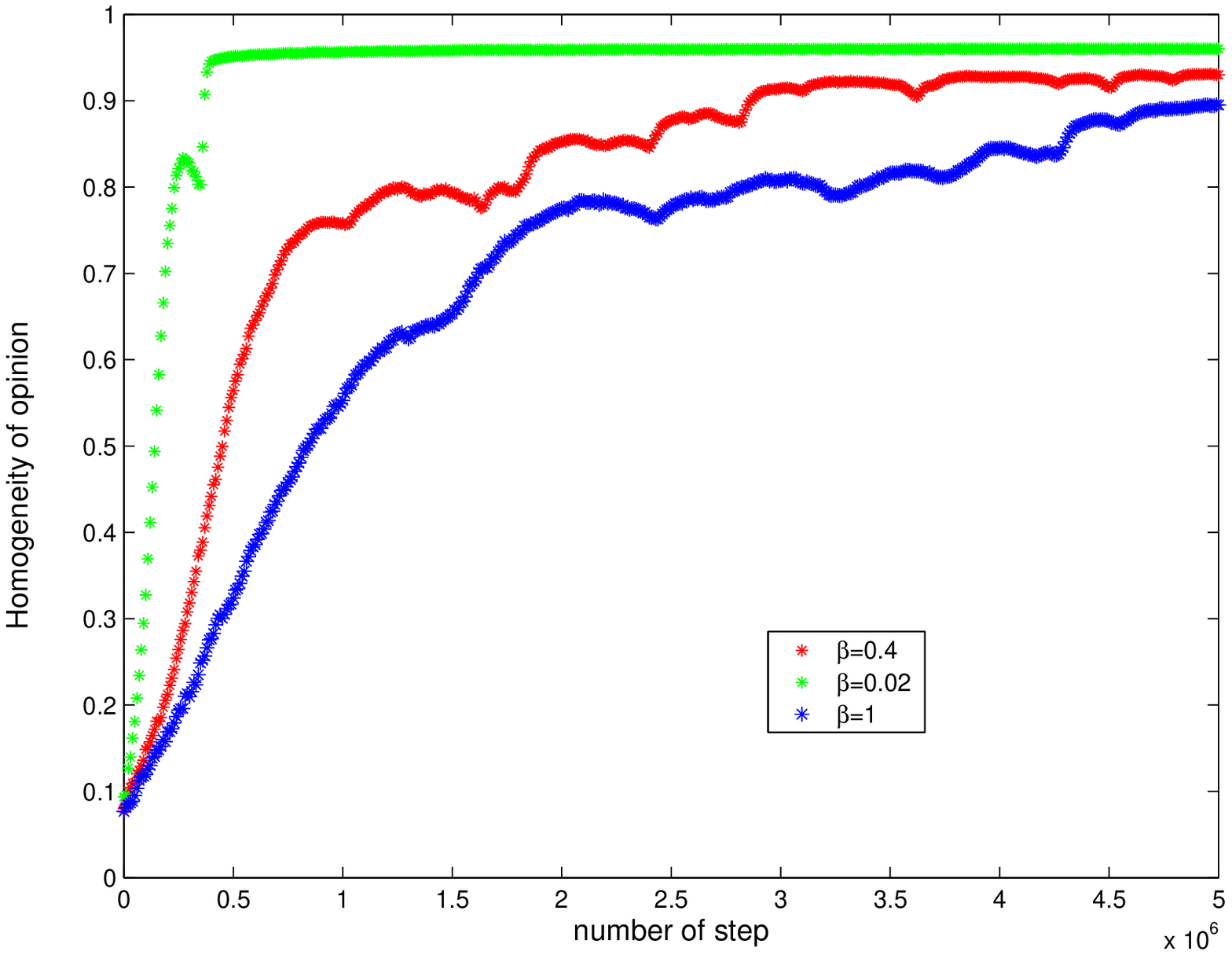,width=3.5TRUEIN}} 
\vspace*{8pt}
\fcaption{SO model with $N=10000$, $\gamma=1/8$, a 
uniform distribution of initial opinions and a uniform distribution 
of the individual characteristic parameters $p$ and $q$. 
The Gini coefficient $G$ is reported as a function of time. 
The three curves correspond to three different values of $\beta$, 
namely, from bottom to top, $\beta=1$, $0.4$, and $0.02$.}
\label{fig4}
\end{figure}
In Fig.~\ref{fig4} we report the Gini coefficient as a function of time 
for three different dynamical evolutions of the SO model, respectively 
corresponding to the cases $\beta=0.02$, $\beta=0.4$ and $\beta=1$. 
These three cases are a good sample of the model behaviours,  
since by definition the parameter $\beta$ is allowed to vary in 
the range [0,1]. 
For each value of $\beta$ considered, the system evolves towards 
a state with a high Gini coefficient. We notice that: 
\begin{enumerate}

\item The qualitative dynamical behaviour does not change with $\beta$. 
 The system reaches an asymptotic value of $G$ which is very close to 1. 
This denotes the presence of a single large cluster (and possibly few small
clusters), for any value of $\beta$. In practice, extreme heterogeneity is 
not allowed in the SO model. 

\item The values of $\beta$ influence the dynamics in two ways: 
A) the stationary value of $G$ is smaller for higher values of  $\beta$; 
B) the converge dynamics is slower for higher values of  $\beta$. 

\end{enumerate}

The existence of a single large cluster in the numerical simulations 
of the SO model is largely due to the fact that we have assumed 
a uniform distribution of the individual characteristic parameters 
$p$ and $q$. Other distributions of the parameters, as for instance 
$F_1(p)$ and $F_2(q)$  gaussian distributed, are equally interesting 
to be investigated. In particualar we expect that the model can give 
different results by tuning mean value and standard deviation of 
the gaussian distribution of characteristic parameters.

\vspace*{1pt}\textlineskip	
\section{Conclusions}		
\vspace*{-0.5pt}
\noindent

In this paper we have shown how strategic game theory can find useful 
in the modeling of opinion formations, as a way to simulate the basic 
interaction mechanisms between two individuals. 
In particular, we have shown how various models of opinion formation 
can be obtained by just changing the rules of the game, i.e. the number 
and the kind of actions an individual can choose from, and also 
the very same characteristics of the individuals. 
In the context of one of the simplest game considered we were 
able to derive, by basic principles, a well known 
model of opinion dynamics, such as the Deffuant et al. model. 
Then, we have generalized the Deffuant et al. model by introducing 
in the game social individuals with two 
characteristic parameters, respectively representing different 
inclinations to change opinion and different abilities in convincing the 
others. Such a game produce the so called Stubborn Individuals and Orators (SO) model. 
We have investigated numerically the dynamics of the SO model in the case 
of all-to-all interactions, and 
in the simplest possible case of a uniform distribution of 
characteristic parameters. In such a case the model converges 
to a single dominant opinion for any value of the control parameters. 
This model can still be generalized by using different distributions 
of characteristic parameters, or by allowing the 
individuals to interact only with the neigbours in a network. 
Many other models can be introduced in the context of strategic game 
theory, so that we hope that our paper can stimulate further research in 
the field of sociophysics.


\nonumsection{References}

\end{document}